\documentclass[default,iicol]{sn-jnl}

\jyear{2022}%

\theoremstyle{thmstyleone}%
\theoremstyle{thmstyletwo}

\theoremstyle{thmstylethree}

\usepackage{amssymb}
\usepackage{hyperref}
\hypersetup{colorlinks=true,citecolor=blue}
\usepackage{multirow}
\usepackage{overpic}
\usepackage{colortbl}
\usepackage{threeparttable}
\usepackage{marvosym}

\newcommand{\rev}[1]{\textcolor{black}{#1}}

\newcommand{\secref}[1]{Sec.~\ref{#1}}
\newcommand{\figref}[1]{Fig.~\ref{#1}}
\newcommand{\tabref}[1]{Tab.~\ref{#1}}
\newcommand{\eqnref}[1]{Eqn.~(\ref{#1})}

\newcommand{\tabincell}[2]{\begin{tabular}{@{}#1@{}}#2\end{tabular}}

\definecolor{mygray1}{gray}{.85}
\definecolor{mygray2}{gray}{.5}
\newcommand{\graytext}[1]{\textcolor{mygray2}{#1}}
\def\etal{{\em et al.}}
\def\ie{\emph{i.e.}}
\def\eg{\emph{e.g.}}

\def\etal{{\em et al.~}}

\def\ourmodel{Polyp-OOD}

\makeatletter
\newcommand{\thickhline}{%
	\noalign {\ifnum 0=`}\fi \hrule height 1pt
	\futurelet \reserved@a \@xhline
}
\makeatother

\begin{document}

\title[OOD-MAE]{\textbf{Rethinking Polyp Segmentation from an Out-of-Distribution Perspective}}

\author[1]{\fnm{Ge-Peng} \sur{Ji}}
\author[1]{\fnm{Jing} \sur{Zhang}}
\author[1]{\fnm{Dylan} \sur{Campbell}}
\author[2]{\fnm{Huan} \sur{Xiong}}
\author[1]{\fnm{Nick} \sur{Barnes}}

\affil[1]{\orgname{Australian National University}, \orgaddress{\city{Canberra}, \country{Australia}}}
\affil[2]{\orgname{Mohamed bin Zayed University of Artificial Intelligence}, \orgaddress{\city{Abu Dhabi}, \country{UAE}}}

\abstract{
Unlike existing fully-supervised approaches, we rethink colorectal polyp segmentation from an out-of-distribution perspective with a simple but effective self-supervised learning approach. We leverage the ability of masked autoencoders -- self-supervised vision transformers trained on a reconstruction task -- to learn in-distribution representations; here, the distribution of healthy colon images. We then perform out-of-distribution reconstruction and inference, with feature space standardisation to align the latent distribution of the diverse abnormal samples with the statistics of the healthy samples.
We generate per-pixel anomaly scores for each image by calculating the difference between the input and reconstructed images and use this signal for out-of-distribution (\ie, polyp) segmentation. 
Experimental results on six benchmarks show that our model has excellent segmentation performance and generalises across datasets. 
Our code is publicly available at \url{https://github.com/GewelsJI/Polyp-OOD}.
}

\keywords{Unsupervised polyp segmentation, Out-of-distribution detection, Masked autoencoder, Abdomen.}

\maketitle

\section{Introduction}

Colorectal cancer is the third leading cause of cancer-related deaths worldwide~\cite{center2009worldwide}. 
The localised stage of colon/rectal cancer has a high 5-year survival rate (91\%/90\%). However, according to the SEER statistics maintained by the American Cancer Society~\cite{seer2023ACS}, the survival rate dramatically decreases at the regional stage (72\%/74\%) and the distant stage (13\%/17\%). From clinical practice, regular colorectal screening is vital for cancer prevention, aiming to find and remove precancerous growths (\eg, abnormal colon or rectum polyps) before they turn malignant. This procedure usually relies on the physicians' experience, and a less experienced physician may fail to identify the precancerous conditions, motivating the need for automatic polyp segmentation techniques.

In the past decades, considerable efforts have been devoted to segmenting colorectal polyps in a data-driven manner, with fully-supervised strategies dominating,
\eg, those that use fully-convolutional networks~\cite{brandao2017fully}, U-shaped models~\cite{jha2019resunetplus}, and attention-based variants~\cite{yeung2021focus,mahmud2021polypsegnet,fan2020pranet}.
Recent works extend the task to spatiotemporal modelling for colonoscopy videos using 3D convolutions~\cite{puyal2020endoscopic} or self-attention modules~\cite{ji2021progressively,ji2022video}. However, they all suffer from some data-related limitations: \textbf{(a)} \textit{insufficiently diverse data}: it is hard to collect diverse positive samples (\ie, those with polyps) since they occur with low-frequency during colonoscopy compared to negative ones; and \textbf{(b)} \textit{expensive labelling}: only experienced physicians can provide ground-truth for the medical images, leading to expensive data annotation.

To alleviate the above limitations, data-efficient learning becomes a potential solution, which harnesses the power of artificial learners with less human supervision, such as semi-supervised~\cite{wu2021collaborative,li2022tccnet,zhao2022semi,zhang2021self} and weakly-supervised~\cite{zhu2023feddm,ruiz2022weakly,dong2019semantic,dong2020weakly} strategies. However, they still require an adequate number of positive samples during training. Further, compared with the fully-supervised setting, semi- or weakly-labelled data can cause more serious model biases.
Alternatively, unsupervised anomaly segmentation solutions hypothesise that a model trained exclusively on normal colonoscopy images can identify anomalous regions when analysing an abnormal sample. Previous methods built a one-class classifier using contrastive learning, where auxiliary pretext tasks (\eg, using synthesised~\cite{tian2021self} and augmented~\cite{tian2021constrained} images) were designed for differentiating normal and abnormal patterns. Their performance relies heavily on an elaborately-designed training pipeline and risks over-fitting on those pseudo-abnormal patterns. 

Reconstruction-based methods can solve such problems by training on medical images in a self-supervised manner, where the essential assumption is that an autoencoder trained to rebuild in-distribution (ID) samples cannot reconstruct out-of-distribution (OOD) samples, \ie, colorectal polyps, as effectively~\cite{chalapathy2019deep}.
However, recent research shows that naive autoencoders can still reconstruct OOD samples with relative low error~\cite{denouden2018improving}, indicating that this framework can not be used directly.
To address this issue, Tian \etal \cite{tian2022unsupervised} presented a memory-augmented self-attention encoder and a multi-level cross-attention decoder based on a masked autoencoder~\cite{he2022masked} with a large masking ratio, aiming to obtain high reconstruction error for the anomalous regions. Instead of complicating a model architecture with a different training pipeline \cite{tian2022unsupervised}, we streamline the training procedure by using negative (healthy) samples and then perform data-adaptive inference to identify anomalous regions.

We argue that, compared with healthy data, medically-abnormal data can be treated as OOD, allowing us to define colorectal polyp segmentation as a per-pixel OOD detection task.
The underlying assumption is that the anomalous regions will have a different distribution compared to the healthy samples.
Following the principles of reconstruction-based detection, we directly use the well-designed training pipeline of MAE~\cite{he2022masked} and serve the reconstruction error to assign an anomaly score.
Then, inference becomes an pixel-wise OOD detection task, allowing us to benefit from the strong distribution-modelling capabilities of MAEs.
However, we find that colorectal polyps vary significantly in appearance, leading to different representations in the latent space.
Directly using MAE is then problematic, because the colorectal polyp features are not compactly distributed, degrading the ability of the network to identify them.
To address this, we propose feature space standardisation to produce a compact but distinctive feature representation for colorectal polyp regions, leading to a simple and effective inference stage.

Our main contributions are
\textbf{(a)} reframing unsupervised polyp segmentation as an out-of-distribution detection task; 
\textbf{(b)} learning a distribution for healthy samples using a masked autoencoder, advantageously \textit{requiring only easily-obtainable healthy samples for training}; and
\textbf{(c)} demonstrating that feature space standardisation improves the network's ability to identify anomalous regions at inference time, and to generalise across datasets.
Our approach exhibits excellent performance at unsupervised polyp segmentation.

\begin{figure*}[t!]
    \centering
    \begin{overpic}[width=0.77\linewidth]{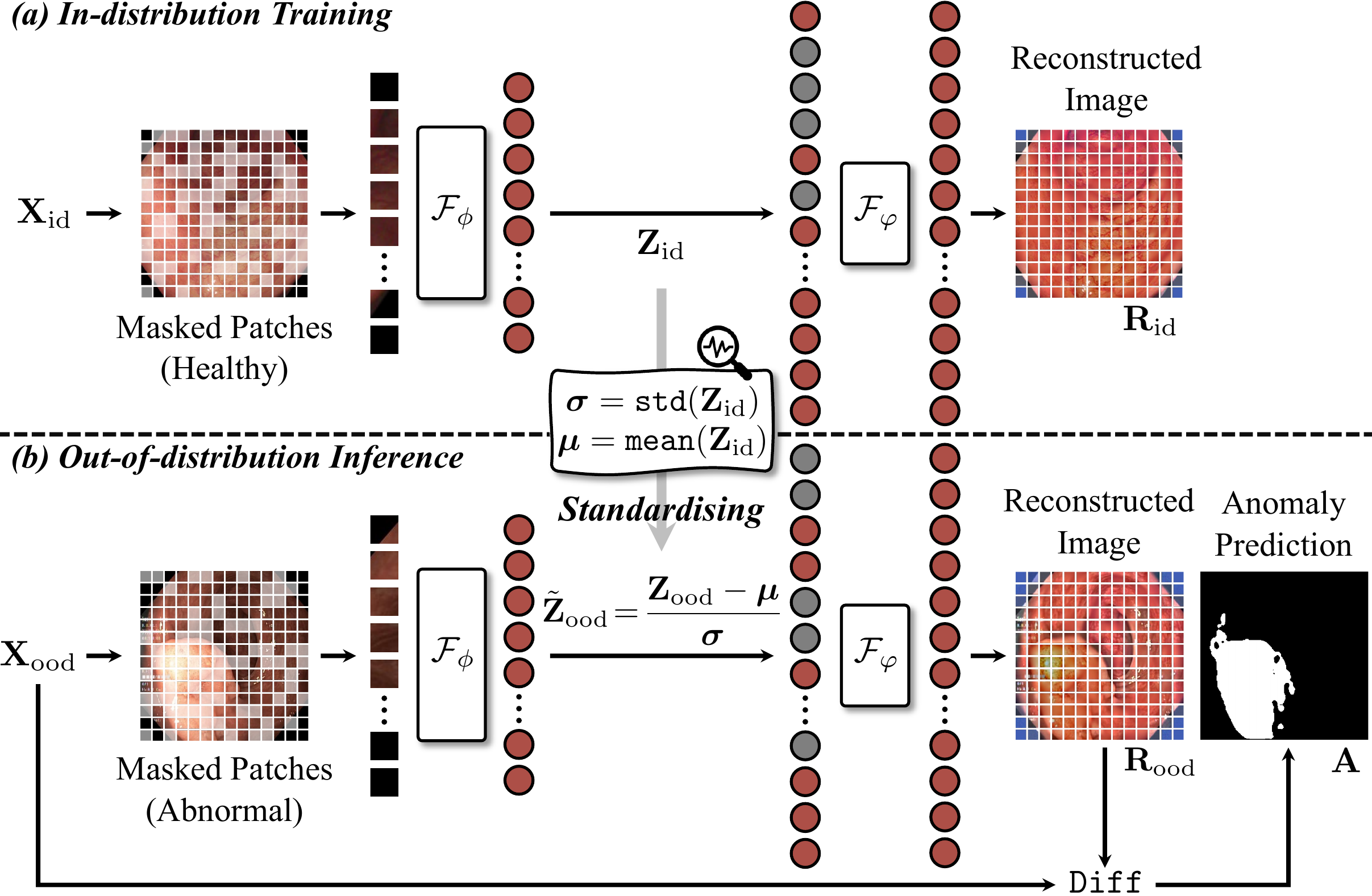}
    \end{overpic}
    \caption{Illustration of our \textbf{\ourmodel} model. (a) ID training: the model is trained to reconstruct the masked-out patches from a healthy sample $\mathbf{X}_{\text{id}}$. (b) OOD inference: we generate an anomaly score $\mathbf{A}$ by calculating the reconstruction difference between abnormal input $\mathbf{X}_{\text{ood}}$ and its reconstruction $\mathbf{R}_{\text{ood}}$.
    }\label{fig:oodmae_framework}
\end{figure*}

\section{Method}

This section describes \ourmodel, our self-supervised learning framework that segments colorectal polyps by exploiting the difference between the in-distribution (ID) and out-of-distribution (OOD) reconstructions of our model. 
The underlying assumption is that a masked autoencoder (MAE~\cite{he2022masked}) trained on healthy samples is capable of learning their distribution and reconstructing healthy regions of the image. At the same time, it should be incapable of modelling and reconstructing OOD regions, such as polyps.
We first revisit MAE in~\secref{sec:revisit_mae}, then present our ID training pipeline in \secref{sec:id_training_pipeline} and our OOD inference pipeline in \secref{sec:ood_inference_pipeline}.

\subsection{Preliminaries: the Masked Autoencoder (MAE)}\label{sec:revisit_mae}
 
An MAE learns a latent representation over the distribution of training images by setting up a self-supervised inpainting target based on partial observations.
Since we leverage MAE training for polyp segmentation, we briefly describe it here.
Interested readers are encouraged to refer to the paper~\cite{he2022masked} for more details.

\noindent$\bullet$~\textbf{Image masking strategy.} Following the data format of ViT~\cite{vit2021ICLR}, an input $\mathbf{X}\!\in\!\mathbb{R}^{C \times H \times W}$ is divided into several non-overlapping patches, where $C$ represents RGB colour channels and $H$/$W$ indicates the image spatial resolution.
Then, according to a randomly generated masking template $\mathbf{M}$ with masking ratio $r$, the patches are labelled as visible or invisible.

\noindent$\bullet$~\textbf{MAE encoder.} Only \textit{visible}, \textit{unmasked} patches are involved at the encoding stage. Specifically, each patch is flattened, linearly projected, and added to a positional embedding before being fed to a transformer encoder $\mathcal{F}_{\phi}$~\cite{vit2021ICLR}, generating a latent representation $\mathbf{Z} = \mathcal{F}_{\phi}(\mathbf{X}; \mathbf{M}; r)$.

\noindent$\bullet$~\textbf{MAE decoder.} The lightweight decoder receives the full token stack, including encoded visible tokens and invisible `empty' tokens\footnote{We replace those invisible, masked-out tokens with a shared learned `empty' vector.} as input. A positional embedding is added to all  tokens to ensure they have valid location information before entering a series of transformer decoding blocks $\mathcal{F}_{\varphi}$. The reconstructed image is synthesised as $\mathbf{R} = \mathcal{F}_{\varphi}(\mathbf{Z})$.

\noindent$\bullet$~\textbf{Optimisation.} The objective of MAE is to `fill' these invisible empty tokens based on an encoded representation from limited visible patches. Specifically, MAE minimises the mean squared error between input $\mathbf{X}$ and output $\mathbf{R}$, and only these masked-out patches are involved in the loss computation.

\subsection{In-distribution Training Pipeline}\label{sec:id_training_pipeline} 

\noindent$\bullet$~\textbf{Healthy dataset preparation.} The first step is to prepare a dataset that contains normal samples from healthy patients. The best candidate is the recently proposed largest-scale colonoscopy video database, SUN-SEG~\cite{ji2022video}, with per-frame dense annotations. It provides 1,106 short videos with 158,690 frames, including positive and negative cases. From this, we collect a dataset of 12,190 healthy images
with a sampling rate of 10 frames. We deliberately do not post-process or remove bad-quality colonoscopy frames, such as motion blur or occlusion, because we want to capture this perturbations in the healthy distribution.

\noindent$\bullet$~\textbf{Training pipeline.} We train an MAE from scratch and explore how it models abnormal regions. As shown in \figref{fig:oodmae_framework}~(a), building upon MAE (see~\secref{sec:revisit_mae}), our model is trained on healthy input $\mathbf{X}_{\text{id}}$, with the forward pass given by
\begin{equation}
    \mathbf{R}_{\text{id}} = \mathcal{F}_{\varphi}(\mathbf{Z}_{\text{id}}) = \mathcal{F}_{\varphi}(\mathcal{F}_{\phi}(\mathbf{X}_{\text{id}}; \mathbf{M}; r)),
\end{equation}
where $\mathbf{Z}_{\text{id}}$ is the ID latent representation.
We apply the loss function to all patches, rather than the visible ones only~\cite{he2022masked}, to optimise for reconstruction.
We observe that with a suitable masking ratio $r$, the reconstruction errors can generalise to distinguishing abnormal polyps and performs well despite not having any human-annotated labels.
We set the masking ratio to $r=35$\% by default, following our empirical analysis in \secref{sec:diagnostic_experiment} and \figref{fig:ablation_mask_ratio}.

\subsection{Out-of-distribution Inference Pipeline}\label{sec:ood_inference_pipeline}

Here we describe how to adapt our ID model to infer abnormal regions.

\noindent$\bullet$~\textbf{Abnormal dataset preparation.} Unlike previous methods that trained and tested on the same colonoscopy dataset~\cite{tian2021constrained,tian2022unsupervised}, we evaluate \ourmodel~in two different settings to better explore the model's generalisability.
First, we form an \textit{in-domain} dataset, called SUN-SEG-I, based on the positive (anomalous) part of SUN-SEG~\cite{ji2022video}.
It is extracted by sampling from 173 positive testing videos every 30 frames, resulting in 563/353 colonoscopy images with easy/hard difficulty levels.
%
Second, following Fan \etal~\cite{fan2020pranet}, we select four popular benchmarks for the \textit{out-of-domain} setting, including the testing data of Kvasir-SEG~\cite{jha2020kvasir}, CVC-ClinicDB~\cite{bernal2015wm}, CVC-ColonDB~\cite{bernal2012towards}, and ETIS-LaribPolypDB~\cite{silva2014toward}.

\noindent$\bullet$~\textbf{Standardised latent space.} Inspired by Han~\etal~\cite{han2022expanding}, we hypothesise that the MAE decoder can identify the high-density regions of latent space better than the low-density regions. In our case, the healthy regions should correspond to the high-density region and the medical anomalies to the low-density region. However, the diverse appearance of polyps may lead to them being mapped to the high-density region as well. To reduce the likelihood of this, we introduce inference-time feature space standardisation, leading to a compact and distinctive feature representation for anomalous regions. 
Specifically, the test sample features are standardised before being decoded back to image space, that is,
\begin{equation}\label{eqn:id_statistic}
    \tilde{\mathbf{Z}}_{\text{ood}} = \frac{\mathbf{Z}_{\text{ood}} - \boldsymbol{\mu}}{\boldsymbol{\sigma}},
\end{equation}
where $\boldsymbol{\mu}=\texttt{mean}(\mathbf{Z}_{\text{id}})$ and $\boldsymbol{\sigma}=\texttt{std}(\mathbf{Z}_{\text{id}})$.
These statistics (mean $\boldsymbol{\mu}$ and standard deviation $\boldsymbol{\sigma}$) have the same dimension as the features $\mathbf{Z}_{\text{ood}}$ and are computed from the entire ID (healthy) dataset.

\noindent$\bullet$~\textbf{Inference pipeline.} As shown in~\figref{fig:oodmae_framework}~(b), given an OOD input $\mathbf{X}_{\text{ood}}$ with abnormal polyps, the reconstructed image $\mathbf{R}_{\text{ood}}\!\in\!\mathbb{R}^{C \times H \times W}$ is generated by
\begin{equation}\label{eqn:ood_inference}
    \mathbf{R}_{\text{ood}} = \mathcal{F}_{\varphi}(\tilde{\mathbf{Z}}_{\text{ood}}) = \mathcal{F}_{\varphi} \left(\frac{\mathcal{F}_{\phi}(\mathbf{X}_{\text{ood}}; \mathbf{M}; r) - \boldsymbol{\mu}}{\boldsymbol{\sigma}} \right),
\end{equation}
where ID descriptive statistics ($\boldsymbol{\mu}$ and $\boldsymbol{\sigma}$) are calculated by~\eqnref{eqn:id_statistic}.
We then compute the pixel-wise errors for all input--reconstruction pairs, and average across the $C$ image channels. The anomaly score $\mathbf{A} \in \mathbb{R}^{1 \times H \times W}$ is predicted as
\begin{equation}\label{eqn:anomaly_score}
\begin{aligned}
\mathbf{A} &= \texttt{Diff}(\mathbf{X}_{\text{ood}}, \mathbf{R}_{\text{ood}}) \\
&= \mathbf{P} \circledast \frac{1}{C} \sum_{c=1}^{C}\Vert \mathbf{R}_{\text{ood}} - \mathbf{X}_{\text{ood}} \Vert_{1},
\end{aligned}
\end{equation}
where $\circledast$ denotes an average-pooling operation $\mathbf{P}$ over the input signal, thus eroding the mask slightly to remove some artefacts.

\begin{table*}[t!]
\centering
\footnotesize
\caption{Quantitative performance on \textit{in-domain}~\cite{ji2022video} and \textit{out-of-domain}~\cite{jha2020kvasir,bernal2015wm,tajbakhsh2015automated,silva2014toward} testing datasets. $\uparrow$ indicates that a higher evaluation score (\%) is better.}
\label{tab:quantitative_comparison}
\renewcommand{\arraystretch}{1.2}
\renewcommand{\tabcolsep}{0.41cm}
\begin{threeparttable}
\begin{tabular}{| r || ccc | ccc | ccc |}
\thickhline
\rowcolor{mygray1}
&\multicolumn{3}{c|}{\tabincell{c}{SUN-SEG-I (Easy)~\cite{ji2022video}}} &\multicolumn{3}{c|}{\tabincell{c}{SUN-SEG-I (Hard)~\cite{ji2022video}}} &\multicolumn{3}{c|}{\tabincell{c}{Kavasir-SEG~\cite{jha2020kvasir}}} \\
\rowcolor{mygray1}
Models &Spe~$\uparrow$ &$\mathcal{S}_\alpha$~$\uparrow$ &$E_{\Phi}$~$\uparrow$ &Spe~$\uparrow$ &$\mathcal{S}_\alpha$~$\uparrow$ &$E_{\Phi}$~$\uparrow$ &Spe~$\uparrow$ &$\mathcal{S}_\alpha$~$\uparrow$ &$E_{\Phi}$~$\uparrow$ \\
\hline
\hline 
SSIM-AE~\cite{bergmann2018improving} &66.8 &35.3 &46.1 &64.0 &34.6 &44.8 &60.7 &32.9 &44.4 \\
IGD~\cite{chen2022deep} &\textbf{81.0} &40.9 &59.3 &76.7 &41.6 &59.9 &87.1 &39.8 &60.6 \\
AnoDDPM~\cite{wyatt2022anoddpm} &71.8 &45.5 &59.1 &73.0 &46.0 &60.8 &91.5 &47.8 &61.2 \\
\textbf{\ourmodel} &76.8 &\textbf{45.9} &\textbf{64.1} &\textbf{80.2} &\textbf{46.3} &\textbf{65.0} &\textbf{93.7} &\textbf{48.2} &\textbf{64.1} \\
\hline
\hline
\thickhline    
\rowcolor{mygray1}
&\multicolumn{3}{c|}{\tabincell{c}{CVC-ClinicDB~\cite{bernal2015wm}}} &\multicolumn{3}{c|}{\tabincell{c}{CVC-ColonDB~\cite{bernal2012towards}}} &\multicolumn{3}{c|}{\tabincell{c}{ETIS-LaribPolypDB~\cite{silva2014toward}}} \\
\rowcolor{mygray1}
Models &Spe~$\uparrow$ &$\mathcal{S}_\alpha$~$\uparrow$ &$E_{\Phi}$~$\uparrow$ &Spe~$\uparrow$ &$\mathcal{S}_\alpha$~$\uparrow$ &$E_{\Phi}$~$\uparrow$ &Spe~$\uparrow$ &$\mathcal{S}_\alpha$~$\uparrow$ &$E_{\Phi}$~$\uparrow$ \\
\hline
\hline
SSIM-AE~\cite{bergmann2018improving} &\textbf{83.4} &39.6 &59.7 &60.0 &36.0 &52.4 &77.1 &39.2 &43.3\\
IGD~\cite{chen2022deep} &76.0 &37.0 &53.8 &69.3 &39.5 &54.2 &75.2 &41.9 &49.3 \\
AnoDDPM~\cite{wyatt2022anoddpm} &73.8 &48.1 &59.6 &64.1 &46.4 &56.9 &72.5 &47.0 &51.2 \\
\textbf{\ourmodel} &77.4 &\textbf{48.5} &\textbf{60.0} &\textbf{72.0} &\textbf{46.7} &\textbf{64.2} &\textbf{81.4} &\textbf{47.0} &\textbf{61.0} \\
\hline
\end{tabular}
\end{threeparttable}
\end{table*}

\begin{figure*}[h!]
    \centering
    \begin{overpic}[width=\linewidth]{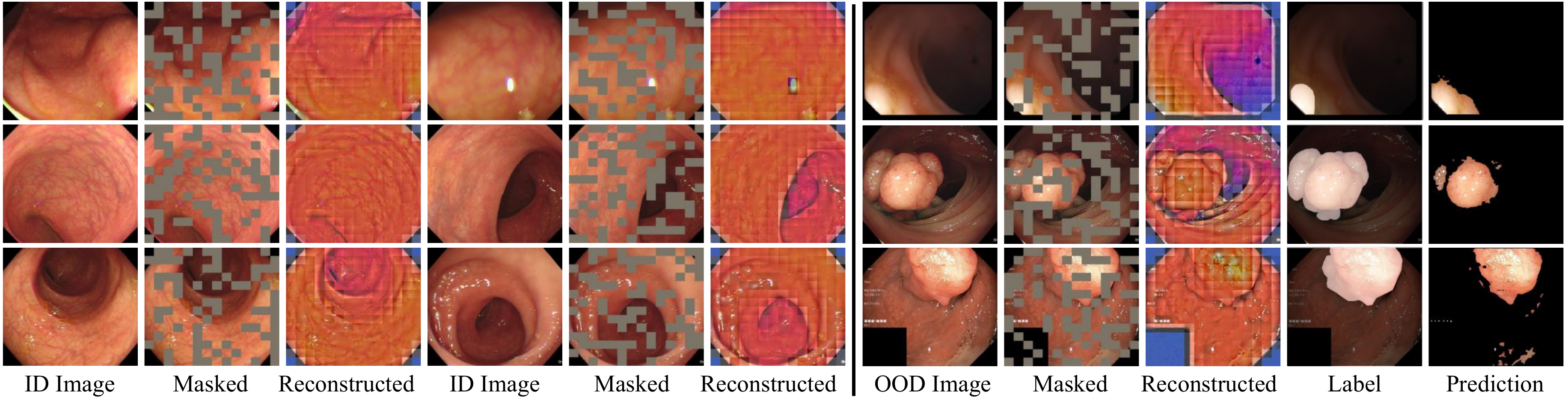}
    \end{overpic}
    \caption{The qualitative results of our \ourmodel~on in-distribution (left part) and out-of-distribution (right part) samples.}\label{fig:qualitative_viz}
\end{figure*}

\section{Experiment}

\subsection{Experiment Setup}\label{sec:experiment_setup}

\noindent$\bullet$~\textbf{Training configuration.} Our model is implemented with the PyTorch framework and trained on three NVIDIA 2080Ti GPUs with 11GB memory. We basically follow MAE's configurations for the pre-training stage: resizing all colonoscopy inputs to a uniform resolution ($H\!=\!W\!=\!224$) with patch size 16, but changing masking ratio to 35\%. We choose ViT-Base/16~\cite{vit2021ICLR} as our transformer encoder and train our model using the AdamW optimiser with batch size 44 per GPU, initial learning rate 1.5e-4, and weight decay 5e-2.

\noindent$\bullet$~\textbf{Evaluation protocols.} For a fair comparison, as reported in~\tabref{tab:quantitative_comparison}, we compare our model with three recent unsupervised medical anomaly segmentation methods~\cite{bergmann2018improving,chen2022deep,wyatt2022anoddpm}. We evaluate performance on three binary segmentation metrics, including structure measure ($\mathcal{S}_\alpha$)~\cite{cheng2021structure}, enhanced-alignment measure ($E_{\Phi}$)~\cite{fan2021cognitive}, and specificity (Spe). In all experiments, we report the maximum value of Spe and $E_{\Phi}$ by iterating over all thresholds from 0 to 255.

\begin{figure*}[t!]
    \centering
    \begin{overpic}[width=0.85\linewidth]{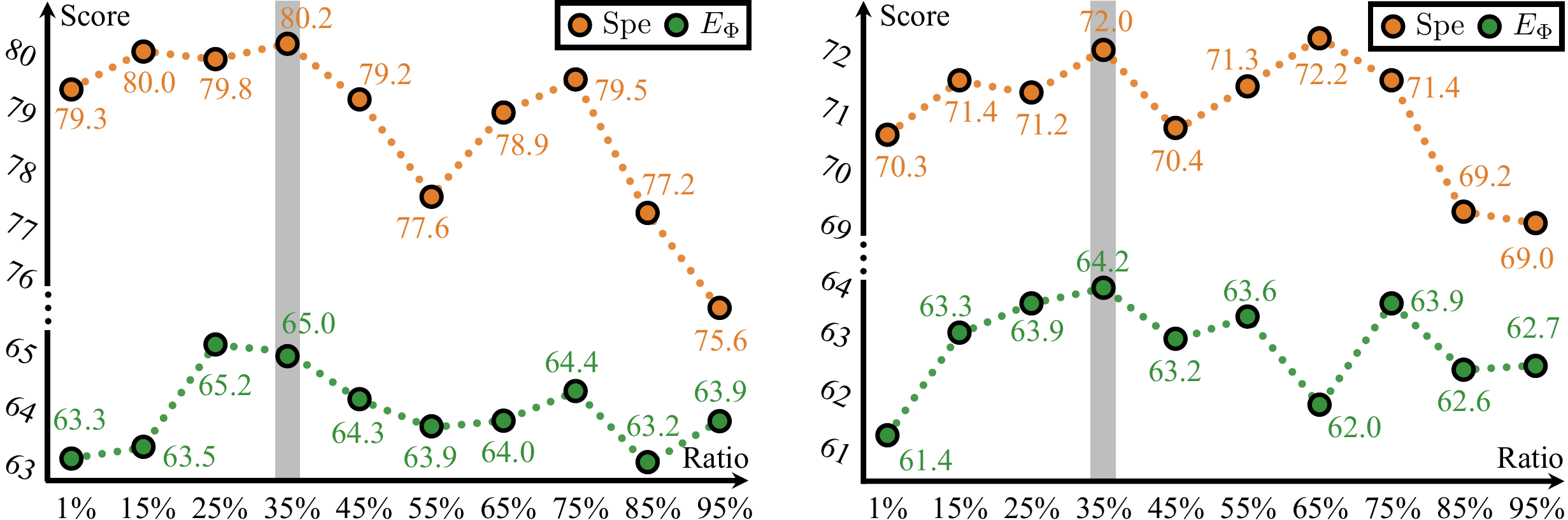}
    \end{overpic}
    \caption{The effect of different masking ratios, including in-domain (Left: SUN-SEG-I (Hard)~\cite{ji2022video}) and out-of-domain (Right: CVC-ColonDB~\cite{bernal2012towards}) datasets.
    }\label{fig:ablation_mask_ratio}
\end{figure*}

\begin{table*}[t!]
\centering
\footnotesize
\caption{Ablation study on standardising the latent space during out-of-distribution inference.}
\label{tab:ablation_standardisation}
\renewcommand{\arraystretch}{1}
\renewcommand{\tabcolsep}{0.56cm}
\begin{threeparttable}
\begin{tabular}{| r || ccc | ccc |}
    \thickhline
    \rowcolor{mygray1}
    &\multicolumn{3}{c|}{\tabincell{c}{SUN-SEG-I (Hard)~\cite{ji2022video}}} &\multicolumn{3}{c|}{\tabincell{c}{CVC-ColonDB~\cite{bernal2012towards}}} \\
    \rowcolor{mygray1}
    Strategy &Spe~$\uparrow$ &$\mathcal{S}_\alpha$~$\uparrow$ &$E_{\Phi}$~$\uparrow$ &Spe~$\uparrow$ &$\mathcal{S}_\alpha$~$\uparrow$ &$E_{\Phi}$~$\uparrow$ \\
    \hline
    \hline
    \textit{w/o} standardisation &73.9 &46.1 &61.4 &62.7 &46.6 &61.9 \\
    \textbf{\textit{w/} standardisation} &\textbf{80.2} &\textbf{46.3} &\textbf{65.0} &\textbf{72.0} &\textbf{46.7} &\textbf{64.2} \\
    performance gain &\graytext{\tiny{(+6.3\%)}} &\graytext{\tiny{(+0.2\%)}} &\graytext{\tiny{(+3.6\%)}} &\graytext{\tiny{(+9.3\%)}} &\graytext{\tiny{(+0.1\%)}} &\graytext{\tiny{(+2.3\%)}} \\
    \hline
\end{tabular}
\end{threeparttable}
\end{table*}

\subsection{In-domain Performance} 

We first validate the model's performance on the \textit{in-domain} dataset, \ie, SUN-SEG-I~\cite{ji2022video} with easy and hard difficulty levels. This setting trains and tests on the same dataset, a homogeneous domain. As reported in~\tabref{tab:quantitative_comparison}, compared with the similarity-based autoencoder (SSIM-AE~\cite{bergmann2018improving}), our model obtains better specificity performance (80.2\% vs. 64.0\%) on the Hard subset of SUN-SEG-I, misdiagnosing healthy patients less frequently.
Our model also has superior performance with respect to the two structure-based metrics ($\mathcal{S}_\alpha$ and $E_\Phi$) on both Easy and Hard subsets, revealing that our predictions have a greater structural similarity with ground-truth.

\subsection{Out-of-domain (Generalisation) Performance} 

As reported in~\tabref{tab:quantitative_comparison}, we also examine the model's ability to generalise to \textit{out-of-domain} datasets, \ie, where the training and testing datasets are different (heterogeneous domains). 
It shows that our model achieves strong results compared with a denoising diffusion probabilistic model (AnoDDPM~\cite{wyatt2022anoddpm}), improving $E_\Phi$ by 7.3\% and 9.8\% on CVC-ColonDB and ETIS-LaribPolypDB, respectively.
Compared with a generative adversarial model (IGD~\cite{chen2022deep}), our model improves $\mathcal{S}_\alpha$ by 8.4\% and 11.5\% on Kvasir-SEG and CVC-ClinicDB, respectively.
This performance improvement validates the superior generalisability of \ourmodel, even on unseen samples from an out-of-domain dataset.

\subsection{Qualitative Visualisation}

\noindent$\bullet$~\textbf{Image reconstruction.} \rev{As shown in left part of \figref{fig:qualitative_viz}, we visualise some reconstructed samples generated by our \ourmodel~on ID images. The results show that our model can better recover the colonic scene, such as depth information and the mucosal vascular textures in the colon. However, as shown in right part of \figref{fig:qualitative_viz}, our model is, by design, unable to reconstruct abnormal polyps for OOD images since it lacks OOD knowledge.}

\noindent$\bullet$~\textbf{Anomaly prediction.} \rev{Additionally, we present segmentation some results in the last column of \figref{fig:qualitative_viz}, predicted by our \ourmodel~model. These predictions indicate that our model can accurately localise the abnormal regions.}

\subsection{Diagnostic Experiment}\label{sec:diagnostic_experiment}

To investigate the effect of core designs, we conduct an ablation study on the in-domain dataset (SUN-SEG-I (Hard)~\cite{ji2022video}) and out-of-domain dataset (CVC-ColonDB~\cite{bernal2012towards}), with default configurations as described in~\secref{sec:experiment_setup}.

\noindent$\bullet$~\textbf{Masking ratio.} \figref{fig:ablation_mask_ratio} shows the effect of using different masking ratios $r$.
As the masking ratio increases, we observe that our models performance peaks with respect to the average of the metrics at $r=35\%$, before reducing again.
Notably, we observe that a relatively small masking ratio is optimal for anomaly prediction. This differs from the optimal masking ratio found by He \etal~\cite{he2022masked} for the image classification task, where a much higher masking ratio performs better. We hypothesise that using a greater proportion of visible patches during training promotes more efficient learning of ID representations for the reconstruction task.

\noindent$\bullet$~\textbf{Standardised latent space.} \tabref{tab:ablation_standardisation} reports the comparison of our model without (1$^{st}$ row) and with (2$^{nd}$ row) the standardisation operation. It shows a clear performance boost for both in-domain (\eg, Spe: +6.3\%, $E_{\phi}$: +3.6\%) and out-of-domain (\eg, Spe: +9.3\%, $E_{\phi}$: +2.3\%) settings. This validates our motivation to use standardisation for rectifying OOD representations with ID statistics in \eqnref{eqn:ood_inference}.

\subsection{Discussion}\label{sec:discussion}

We observe `gridding' artefacts in the reconstructed images, as shown in~\figref{fig:qualitative_viz}. We attribute this to the low input resolution (224) and the large patch size (16). With greater computational resources, which permits higher resolution inputs and smaller patches, we expect performance improvements for the polyp segmentation task. \rev{Therefore, it warrants exploring this further by employing high-fidelity synthesis techniques on MAE, such as those discussed in~\cite{li2023mage}.}

\section{Conclusion}

This paper reframes unsupervised polyp segmentation as an out-of-distribution detection task. We propose a simple-to-implement that performs well. It learns the distribution of healthy image patches and uses this to identify anomalous regions at inference time. We observe the critical importance of aligning the latent space at inference time by a standardisation operation that promotes OOD identification and generalisation to out-of-domain datasets.

\section*{Acknowledgments}
The authors would like to thank the anonymous reviewers and editor for their helpful comments on this manuscript.

\section*{Conflicts of Interests}
The authors declared that they have no conflicts of interest in this work. We declare that we do not have any commercial or associative interest that represents a conflict of interest in connection with the work submitted.

\bibliographystyle{IEEEtran}
\bibliography{mir-article}







\end{document}